\documentclass[prl,twocolumn,superscriptaddress]{revtex4-1}
\pdfoutput=1
\usepackage{amsfonts,amsmath,amssymb}
\usepackage{enumerate}
\usepackage{hyperref}
\usepackage{graphicx}
\newcommand{\be}{\begin{equation}}
\newcommand{\ee}{\end{equation}}

\newcommand{\bea}{\begin{eqnarray}}
\newcommand{\eea}{\end{eqnarray}}

\newcommand{\bal}{\begin{align}}
\newcommand{\eal}{\end{align}}

\renewcommand{\d}{{\rm d}}

\newcommand{\bes}{\begin{subequations}}
\newcommand{\ees}{\end{subequations}}

\newcommand{\nn}{\nonumber}

\newcommand{\ra}{\rightarrow}

\begin{document}

\title{Holography of Wrapped M5-branes and Chern-Simons theory}
\author{Dongmin Gang}
\affiliation{School of Physics, Korea Institute for Advanced Study, Seoul 130-722, Korea}
\author{Nakwoo Kim}
\affiliation{Department of Physics
and Research Institute of Basic Science,  Kyung Hee University, 
 Seoul 130-701,
  Korea}
\author{Sangmin Lee}
\affiliation{Center for Theoretical Physics and
College of Liberal Studies, Seoul National University,
Seoul 151-742, Korea}
\affiliation{School of Physics, Korea Institute for Advanced Study, Seoul 130-722, Korea}

\begin{abstract}
We study three-dimensional superconformal field theories on wrapped M5-branes.
Applying the 
gauge/gravity duality and the 
recently proposed 3d-3d relation, we deduce 
quantitative predictions for the perturbative free energy of a Chern-Simons theory on 
hyperbolic 3-space. Remarkably, the perturbative expansion is expected to terminate at 
two-loops in the large $N$ limit. We check the correspondence numerically in a number of 
examples, and confirm the $N^3$ scaling with precise coefficients.
\end{abstract}

\maketitle

{\it Introduction. }
In quantum field theories, duality refers to a map between observables 
of two seemingly unrelated theories. 
Duality can be particularly powerful 
when one of the two theories is not (yet) defined rigorously. 
There are two prominent examples in string theory: M-theory and  holographic gauge/gravity duality \cite{Maldacena:1997re}. 
While less well-understood than perturbative string theory, 
M-theory offers a unifying framework for all string theories. The gauge/gravity duality relates a quantum field theory  to a quantum gravity theory in one higher dimensions. Although the gravity theory operates mostly at the classical level, it often gives powerful predictions for the quantum field theory.  

A number of new dualities have been discovered recently through
compactification of M5-branes. 
Just as M-theory unifies 
string theories, M5-branes provide a unifying framework for a large class of 
supersymmetric quantum field theories. In the simplest case, the M5-brane theory defines a 6d conformal field theory with $(2,0)$ supersymmetry. Wrapping M5-branes 
on  internal manifolds gives rise to lower dimensional field theories 
with the same or a smaller number of supersymmetries.

In conventional compactifications, the compact manifold affects the definition of
the lower dimensional field theory, 
but does not usually bear an independent physical meaning. 
A novelty in recent works on M5-branes is that a duality holds between the compactified field theory and a different field theory defined {\em on the internal manifold}. For instance, in the celebrated ``4d-2d" relation \cite{Gaiotto:2009we,*Alday:2009aq,*Wyllard:2009hg} 
a 4d $\mathcal{N}=2$ supersymmetric field theory is paired with 
an integrable field theory on a Riemann surface. 
Similarly, the ``3d-3d" relation \cite{Terashima:2011qi,*Terashima:2011xe,*Dimofte:2011ju,*Dimofte:2011py,*Gang:2013sqa,*Yagi:2013fda,*Lee:2013ida,*Cordova:2013cea} 
connects a 3d $\mathcal{N}=2$ supersymmetric field theory with
a 3d Chern-Simons (CS) theory.

The goal of this Letter is to point out and verify a surprising prediction for 
the perturbative expansion of CS theory, which is deduced from 
a combination of the gauge/gravity duality and the 3d-3d relation.
We report on the main results here, and the details
will be published elsewhere \cite{Gang:2014b}.  

We begin with wrapping a stack of $N$ M5-branes on a hyperbolic 3-manifold 
$M$. The resulting lower dimensional theory is called $T_N[M]$ \cite{Dimofte:2013iv}. One of the fundamental observables of the theory is the partition function 
on a squashed three-sphere, $Z_{T_N[M]}[S^3_b]$, with a squashing parameter $b$, and the associated free energy $F_{N,b} = - \log |Z_{T_N[M]}[S^3_b]|$. 
We will use the dualities to study properties of $F_{N,b}$ without computing it directly from $T_N[M]$.  

On the one hand, we embed the brane configuration into the full M-theory 
to invoke the gauge/gravity duality. Building upon the relevant supergravity solution \cite{Gauntlett:2000ng,
*Donos:2010ax} and taking the squashing into account \cite{Martelli:2011fu}, we will show that 
the gravity computation gives  
$F^{\rm gravity} = N^3(b+b^{-1})^2 {\rm vol}(M)/12\pi$ in the large $N$ limit.  
Gauge/gravity duality leads to an equality between the gravity free energy and field theory free energy $F_{N,b}$ at large $N$. 
On the other hand, we use the 3d-3d relation to compute $F_{N, b}$ 
from the CS theory. The methods for the computation 
were developed recently in \cite{Dimofte:2011gm,Dimofte:2013iv}. 
A crucial feature of the 3d-3d relation is that the 
loop-counting parameter ``$\hbar$" of the perturbative CS theory 
is related to the squashing parameter $b$ 
as $\hbar = 2\pi i b^2$ \cite{Terashima:2011qi,Beem:2012mb}. It follows that the  $n$-th term $F^{(n)}_N$, 
defined as
\begin{align}  
F^{\rm CS}_{N,b} = \sum_{n=0}^\infty (\hbar/i)^{n-1} F_N^{(n)} +(\textrm{non-perturbative})\,, 
\label{perturbative-expansion}
\end{align}
comes from the $n$-loop diagrams of the perturbative CS theory. 
Comparing this asymptotic expansion with the gravity free energy, we infer: 
 (1) $F^{(0)}_N,F^{(1)}_N$  and $F^{(2)}_N$ all scale as $N^3$  and their coefficients of $N^3$ are proportional to ${\rm vol}(M)$.
(2) Three- and higher-loop terms as well as  the non-perturbative ones 
are suppressed at large $N$. 

After reviewing the gravity computation and the methods 
for the CS computation, we subject our main observation to numerical tests. 
For a number of hyperbolic knot complements, and the value of $N$ reaching up to 30, our numerical results exhibit excellent  agreement with 
the predictions of the dualities.

{\it Supergravity description. }
It is convenient to use lower dimensional gauged supergravity for constructing various
 near-horizon geometries of D- or M-brane backgrounds. For M5-branes the relevant 
 theory is 7d $SO(5)$ gauged supergravity, which is a consistent truncation 
of 11d supergravity. In addition to the maximally supersymmetric $AdS_7$,
it exhibits a rich spectrum of magnetically charged $AdS$ solutions which we  interpret as 
M5-branes wrapped on supersymmetric cycles \cite{Gauntlett:2000ng}. 

In particular, we are interested in an $AdS_4\times M$ solution where
M5-branes are wrapped on a special Lagrangian 3-cycle $M$ which is locally $H^3$, the hyperbolic 3-space.
To implement topological twisting, one first turns on $SO(3)\subset SO(5)$ part of gauge fields so that
they exactly cancel the contribution of spin connections on $H^3$ in the Killing spinor equation.
There are also 14 scalar fields in the traceless symmetric tensor representation of $SO(5)$,
and we turn on a single scalar field
which is singlet under the remaining symmetry $SO(3)\times SO(2)$.  

It turns out that the supersymmetry and the equation of motion uniquely determine the $AdS_4\times H^3$ solution \cite{Gauntlett:2000ng}. 
One can then use the uplifting formula to obtain a solution of 11d
supergravity. The metric is 
\begin{align}
&\d s^2_{11} = \frac{2^{2/3} (1+\sin^2\theta)^{1/3}}{g^2}\Bigg[ \d s^2(AdS_4) + \d s^2(M) 
\nn\\
& \qquad +\frac{1}2 \left(d\theta^2 + \frac{\sin^2\theta}{1+\sin^2\theta} d\phi^2 \right) + \frac{\cos^2\theta}{1+\sin^2\theta} d\tilde \Omega_2
\Bigg] ,
\label{metric}
\end{align}
where $0<\theta<\pi/2, 0<\phi<2\pi$. $M$ is locally $H^3$. 
Both $AdS_4$ and $M$ have unit radius.
$\d \tilde \Omega_2$ denotes the unit 2-sphere, 
twisted by the spin connection one-forms 
of $M$.


The parameter $g$ is the coupling constant of 7d supergravity, and 
sets the overall curvature scale of the solution. Through the flux quantization, $g$
is related to the number of M5-branes $N$.
The 4-form field $G$ of 11d supergravity, when restricted to the internal space $X_4$, is
\be
G |_{X_4}= -\frac{1}{g^3}\d\left[
\frac{\cos^3\theta}{1+\sin^2\theta}
\right]
 \wedge \d\phi \wedge {\rm vol} (\tilde S^2) \, . 
\label{gflux}
\ee
Integrating this, one obtains
$N = (\pi l_{\rm P}^3 g^3)^{-1}$, 
where $l_{\rm P}$ is 11d Planck length.

The gravity side computation of the partition function can be done using
the standard AdS/CFT prescription. That is, we calculate the holographically renormalized
on-shell action for the supergravity solution. For round $S^3$, the result is simply 
$F = \frac{\pi}{2G_4}$, where $G_4$ is 4d Newton's constant. See e.g. \cite{Emparan:1999pm,
*Gabella:2011sg} for derivation. 

To invoke the 3d-3d relation we put the wrapped M5-brane theory on an ellipsoid
$S^3_b$, defined by $b^2 (x_1^2+x_2^2)+b^{-2}(x_3^2+x_4^2)=1$. 
The geometry has a manifest $b\leftrightarrow b^{-1}$ symmetry and so do all partition functions in this Letter.   
For the holographic computation on $S^3_b$, we consider the 
minimal ${\cal N}=2$ gauged supergravity in 4d, and look for a particular supersymmetric solution 
whose metric and the Killing spinors reproduce the 
$S^3_b$ metric and its Killing spinor given in \cite{Hama:2011ea}, as one approaches the boundary. Such a solution is presented in \cite{Martelli:2011fu}, which is a class of Plebanski-Demianski 
solutions in Einstein-Maxwell theory. Then the 11d solution \eqref{metric} should  change accordingly, 
as one plugs the solution in \cite{Martelli:2011fu} into the uplifting formula of \cite{Gauntlett:2000ng}.
But it is also established in \cite{Martelli:2011fu} that  the $b$-dependence of the holographic free energy is universally given as $F_b = (b+1/b)^2F_{b=1}/4$. One may thus first compute $F_{b=1}$
using \eqref{metric} and restore $b$-dependence easily.
\be
F^{\rm gravity} = \frac{N^3}{12\pi}\left( b+\frac{1}{b} \right)^2 \textrm{vol}(M) \, .
\label{gravity free energy}
\ee
This is the key result we check against the field theory in this Letter. 
Since the gravity analysis is classical, $F^{\rm gravity}$ 
captures only the leading $N^3$ term at large $N$. On the other hand, its $b$-dependence is exact
as coefficient of $N^3$. For knot complements $M = S^3\backslash K$, the solution \eqref{metric} needs to be modified to incorporate intersecting M5-branes along the knot.  For 4d theories of class $S$ associated with a Riemann surface $\Sigma_{g,h}$ of genus $g$ with $h$ full punctures, the leading $N^3$ terms of conformal anomaly coefficients $a$ and $c$ depend only on the Euler characteristic of the Riemann surface regardless of the existence of punctures \cite{Gaiotto:2009gz}. In a similar vein, as the hyperbolic volume is a topological invariant, we expect the formula \eqref{gravity free energy} to be robust and insensitive to the presence of the knot $K$.

{\it 3d-3d relation and a $PGL(N)$ CS theory. }
The 3d-3d relation \cite{Terashima:2011qi,Dimofte:2011ju,Cordova:2013cea} 
states a precise map between $T_N[M]$ and the analytically continued $PGL(N)$ CS theory on $M$. The map for supersymmetric partition function is 
\bea
&Z_{T_N[M]}[S^3_b]= Z^{\rm CS}_{N}[M;\hbar]\;.  \label{dictionary 1}
\eea
In this Letter, we focus on the case when  the 3-manifolds are  hyperbolic knot complements  on $S^3$, $M=S^3\backslash K$, obtained by removing a tubular neighborhood of a hyperbolic knot $K$ from $S^3$. 
A unique complete hyperbolic metric is known to exist for each $M=S^3\backslash K$. 
For the notation of knots we follow
\cite{rolfsen:knots}. 
The volume of $M$ can be expressed in terms of dilogarithm, e.g. $\textrm{vol}(S^3\backslash {\bf 4}_1) = 2\textrm{Im}\big{(}{\rm Li}_2 (e^{\frac{i\pi} 3}) \big{)}= 2.02988
\cdots$. 

 A knot complement $M$ has a torus boundary and  $T_{N}[M]$ has
a flavor symmetry of rank $N-1$ which will be enhanced to $SU(N)$ at IR \cite{Dimofte:2013iv}. 
Both sides of \eqref{dictionary 1} 
are functions of complex parameters $\{\mu_i\}_{i=1}^{N-1}$. 
For $T_{N}[M]$, $\mu_i$ are complexified mass parameters 
 \bea
\mu_i=2 \pi b \left( m_i+ \frac{i}2 (b + b^{-1}) r_i \right) \;,  \label{dictionary 3}
\eea
where $m_i$ and $r_i$ are real masses  and R-charges coupled to the $U(1)^{N-1}$ flavor symmetry. 
For comparison with $AdS_4$ gravity, the conformal symmetry requires $m_i = 0$ and $r_i$ are 
determined via maximization of the free energy on $S^3$  \cite{Jafferis:2010un}. The symmetry enhancement
to $SU(N)$ leads to $r_i = 0$ which are invariant under Weyl reflections. For the CS theory, we consider a boundary condition  which fixes the conjugacy class of gauge holonomy along the meridian cycle of $\partial M$. $\mu_i$ parametrizes the meridian  holonomy.

The action  for the CS theory is 
\bea
&&S_{\rm CS}[{\cal A},\bar{{\cal A}}]= \frac{i}{2 \hbar} CS[\mathcal{A}] + \frac{i}{2 \tilde{\hbar}}CS[\bar{\mathcal{A}}]\;, 
\\
&&CS[\mathcal{A}]:=\int_M \textrm{Tr}({\cal A}\wedge d {\cal A} + \frac{2}3 {\cal A} \wedge {\cal A} \wedge{\cal A})\;.
\eea
We  consider an analytic continuation of the  theory \cite{Witten:2010cx} where  $\hbar, \tilde{\hbar}$ are  complex and ${\cal A},\bar{{\cal A}}$ are  independent gauge fields. 
$\hbar$ and $\tilde{\hbar}$ are mapped through the 3d-3d relation  to the squashing parameter $b$ as \cite{Terashima:2011qi,Beem:2012mb} 
\bea
\hbar = 2\pi i b^2\;,\quad \tilde{\hbar} = - 4\pi^2/\hbar =2\pi i b^{-2}\;. \label{dictionary 2}
\eea
Formally, $Z_{N}^{\rm CS}[M]$ can be written as a path-integral,
\begin{align}
Z_{N}^{\rm CS}[M](\mu_i) = \int D{\cal A}D\bar{{\cal A}}|_{\rm b.c.} e^{i S_{\rm CS}[{\cal A},\bar{\cal A}]}\;, \label{path-integral}
\end{align}
with the boundary condition $|_{\rm b.c.}$ specified by $\{\mu_i\}$. 
In practice, it is more convenient to use canonical quantization. 
The classical solutions are flat-connections, 
\bea
\mathcal{F} = d{\cal A}+{\cal A}\wedge {\cal A}=0\;, \quad 
\bar{\mathcal{F}} = d{\bar{\cal A}}+{\bar{\cal A}}\wedge {\bar{\cal A}}=0\;.
\eea
For quantization, we first consider a classical phase space ${\mathcal{P}(\partial M)}$  
associated with the boundary of $M$, 
\bea
&\mathcal{P}_{N}(\partial M) = \{\mathcal{A},\bar{\mathcal{A}} | \mathcal{F} = \bar{\mathcal{F}} = 0  \textrm{ on $\partial M$}\}/\mbox{(gauge)} \;,\nn
\eea
and  its  Lagrangian submanifold associated with $M$  \cite{Dimofte:2009yn},
\bea
&\mathcal{L}_N(M) = \{\mathcal{A},\bar{\mathcal{A}} | \mathcal{F} = \bar{\mathcal{F}} = 0  \textrm{ on $M$}\}/\mbox{(gauge)} \;.  \nn
\label{classical datum}
\eea
After quantization, the  phase space  is replaced by  a Hilbert-space $\mathcal{H}_{N}(\partial M)$, and $\mathcal{L}_{N}(M) $ by a state $|M_N \rangle \in \mathcal{H}_{N}(\partial M)$. 
The dimension of the phase space is $2(N-1)$ and we choose the  meridian $\{\mu_i\} $ as position variables. Collecting all the ingredients, the CS partition function \eqref{path-integral} can be identified as a  wave-function \cite{Dimofte:2009yn},
\begin{align}
Z^{\rm CS}_{N} [M] (\mu_i) = \langle \mu_i|M_N\rangle\;. \label{CS ptn in canonical quantization}
\end{align}

It is possible to write down an
integral expression for $Z_N^{\rm CS}$, thanks to the two recently developed tools: $N$-decomposition of  $M$ \cite{Dimofte:2013iv} and  a state-integral model in \cite{Dimofte:2011gm}.  
They both make use of an ideal triangulation of $M$,
\bea
M = \left( \bigcup_{i=1}^k   \Delta_i \right)/(\textrm{gluing data})\;.
\eea
\begin{figure}[h!]    
\begin{center}
   \includegraphics[width=0.4\textwidth]{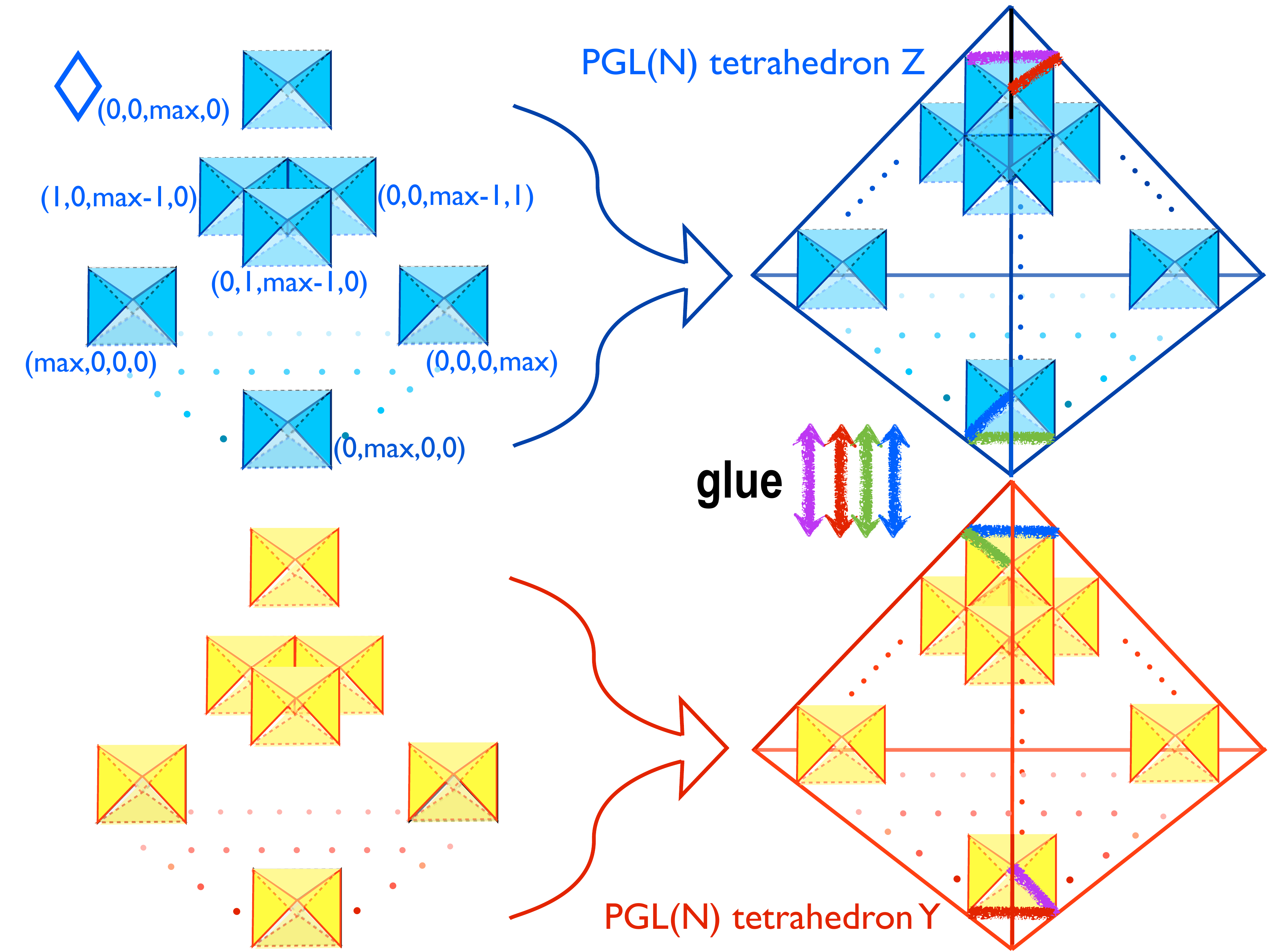} 
   \end{center}
   \caption{$N$-decomposition for $M=S^3\backslash \mathbf{4}_1$. $M$ is decomposed into  two tetrahedra $Y$ and $Z$. Each tetrahedron is decomposed into a pyramid of $\tfrac{1}{6} N (N^2-1)$ octahedra.  }
    \label{fig:N-decomposition for 41}
\end{figure}

\noindent
Dividing each $\Delta_i$ further into a pyramid of $N (N^2-1)/6$ octahedra ($\Diamond$), we obtain a $N$-decomposition of $M$, 
\bea
M  = \left( \bigcup_{i=1}^k   \bigcup_{(a,b,c,d)} \Diamond^{(i)}_{(a,b,c,d)} \right) /(\textrm{gluing data})\;.
\eea
The gluing data dictate how we should match the vertices from different octahedra. 
The octahedra in each $\Delta_i$ are labelled by four non-negative integers $(a,b,c,d)$ whose sum is $N-2$. 
 The decomposition is a mathematical tool to construct the moduli spaces ${\cal P}(\partial M)$ and ${\cal L}(M)$ by `gluing' the building blocks ${\cal P}(\partial \Diamond)$ and ${\cal L}(\Diamond)$ . 

The  state-integral model \cite{Dimofte:2011gm,2012arXiv1202.6268D} is obtained by quantizing the gluing procedure of constructing flat connection moduli spaces.    The model provides a finite dimensional integral expression for $Z^{\rm CS}_N[M]$. 
At conformal point $\mu_i=0$,  ($\mathcal{M}_N:= \frac{k}6 N(N^2-1)$)
\begin{align}
&Z^{\rm CS}_{N}[M]=  \frac{1}{\sqrt{\det B_N}}\int \frac{d^{\mathcal{M}_N} X}{(2\pi  \hbar)^{\mathcal{M}_N/2}}    \prod \psi_{\hbar}(X) \times 
\nn \\
& \exp \left[ - \frac{1}{\hbar}(i \pi
+\frac{\hbar}2) X^T B_N^{-1} \nu_N + \frac{1}{2\hbar } X^T B_N^{-1} A_N   X \right] ,
\label{CS ptn at conformal point}
\end{align}
up to prefactors independent of $N$ and an overall phase factor.
$\psi_{\hbar}(X)$ is a non-compact quantum dilogarithm function, 
which is roughly $Z^{\rm CS}_2[\Diamond]$ \cite{Dimofte:2011gm}. 
$\{A_N, B_N\}$  are  $\mathcal{M}_N \times \mathcal{M}_N$ matrices  and $\nu_N$ is an  $\mathcal{M}_N$-dimensional column vector  with  integer entries. They  can be determined  from the gluing data of the $N$-decomposition up to a certain ambiguity which does not affect our discussion. 

{\it Perturbative CS theory vs gravity.  }
In the limit $\hbar\rightarrow 0$, $Z^{\rm CS}_N[M]$ can be evaluated perturbatively using the saddle point approximation leading to an expansion of the form \eqref{perturbative-expansion}. 
The perturbative ``invariants" $F_N^{(n)}$ can be systematically computed using the Feynman rules 
derived in \cite{2012arXiv1202.6268D}. 
Remarkably enough, in view of the dictionary \eqref{dictionary 2}, we find the gravity free energy \eqref{gravity free energy} displays the same expansion structure as the CS counterpart but terminates at two-loop. Combining 
$F^{\rm gravity} = F^{\rm gauge}$ with the 3d-3d relations \eqref{dictionary 1} and \eqref{dictionary 2}, we 
conclude%
\begin{align}
&\lim_{N \rightarrow \infty }\frac{F_{N}^{(n)}}{N^3}  = c_n  \textrm{vol}(M)\;,     \label{conjecture}
\end{align}
with $c_0=\frac{1}6, c_1=\frac{1}{6\pi}, c_2=\frac{1}{24\pi^2}$ and $c_n=0$ for $n\geq 3$. If the predictions are correct, the symmetry $b\leftrightarrow b^{-1}$
 exists even in the perturbative expansion at large $N$, which gives a strong evidence that   non-perturbative corrections in \eqref{perturbative-expansion}  will be suppressed in the limit.  
The prediction on the classical part $F^{(0)}$ can be understood intuitively \cite{Dimofte:2013iv}. First, we recall that ${\rm Im}(CS[{\cal A}])$ for $PGL(2)$ is equivalent to the 3d $AdS$ gravity action \cite{Witten:1988hc}. 
The unique complete hyperbolic metric on $M$ is mapped to a geometrical flat connection ${\cal A}^{(\rm geom)}_{N=2}$.
The  flat $PGL(2)$-connection can be lifted to  a flat $PGL(N)$-connection ${\cal A}_N^{({\rm geom})}$ using 
the irreducible $N$-dimensional representation of $PGL(2)$.
We assume that  conjugate  $\overline{{\cal A}_N^{({\rm geom})}} $ of   the $PGL(N)$  gives
 a dominant contribution to the path-integral \eqref{path-integral} when $\mu_i=0$. Elementary algebra gives  
 \bea
 CS[\overline{{\cal A}^{({\rm geom})}_N}] =\frac{1}6 N (N^2-1) CS[\overline{{\cal A}^{({\rm geom})}_2}]\;. \label{embedding of PGL(2) to PGL(N)} 
 \eea
Combining this with the fact  that $F_2^{(0)} = {\rm Im}(CS)/2$ for $\overline{{\cal A}^{(\rm geom)}_{N=2}}$ 
equals to $\textrm{vol}(M)$, we arrive at \eqref{conjecture} for $n=0$. The prediction on $F^{(1)}_N$ can   be  proved using results  in \cite{2011arXiv1110.3718M}. A  perturbative analysis of the CS theory gives   $F^{(1)}_N =-\frac{1}2 \log |\textrm{Tor}_{\textbf{adj}}[M, \overline{\mathcal{A}^{\textrm{(geom)}}_N}]|$ where $\textrm{Tor}_{\rho}[M, \mathcal{A}]$ is the Ray-Singer torsion  of an associated vector bundle in a representation $\rho$ twisted by a flat connection $\mathcal{A}$. In \cite{2011arXiv1110.3718M}, it is proven that 
\begin{align}
\lim_{m\rightarrow \infty}\frac{1}{m^2} \log \textrm{Tor}_{\rho_m}[M, \mathcal{A}^{(geom)}_{N=2}]= - \frac{1}{4\pi} \textrm{vol}(M)\;, 
\end{align}
where $\rho_m$ is the irreducible $m$-dimensional representation of $PGL(2)$. Applying the theorem to $F^{(1)}_N$ using the branching rule $\textbf{adj}= \rho_{3}\oplus \rho_{5} \oplus \ldots \oplus \rho_{2N-1}$, we arrive at \eqref{conjecture} for $n=1$.

We currently have little analytic understanding of the loop invariants $F_N^{(n)}$ $(n \ge 2)$. In particular, 
the appearance of ${\rm vol}(M)$ in the 2-loop term is striking and seems non-trivial to prove.

\begin{figure}[h!]    
\begin{center}
   \includegraphics[width=0.5\textwidth]{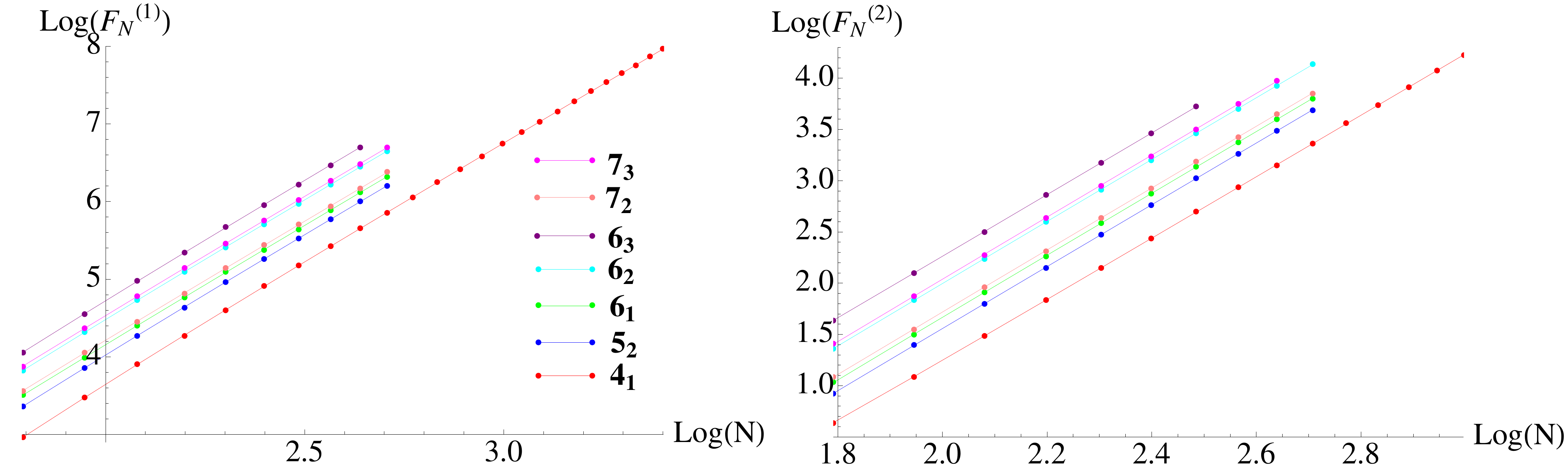} 
   \end{center}
   \caption{Log-log plot of $F_N^{(1)}$ (Left) and  $F_N^{(2)}$(Right) vs. $N$,  for  $N=6,\cdots,N_{\rm max}$ for the seven simplest hyperbolic knot complements $M=S^3\backslash K$ ($K={\bf 4}_1,{\bf 5}_2,{\bf 6}_1,{\bf 6}_2,{\bf 6}_3,{\bf 7}_2,{\bf 7}_3$). $N_{\rm max}$ for each $M$ is limited by computing time.}
    \label{fig:Log-Log}
\end{figure}

We  have verified \eqref{conjecture} for several 
examples of $M$ by calculating the invariants $F_N^{(1)}$, $F_N^{(2)}$ and $F_N^{(3)}$ numerically as we vary $N$. 
The computation of the gluing data $\{A_N,B_N, \nu_N\}$ is greatly facilitated by the computer package 
{\tt SnapPy} \cite{snappy,2012arXiv1207.6711G}. Our results are summarized 
in Fig. \ref{fig:Log-Log}, which shows log-log plots of $F^{(1)}$ and $F^{(2)}$.
They clearly exhibit the expected $N^3$ behavior already at modest values of $N\sim 10$.

To extract the coefficient of $N^3$ term efficiently, we computed the third-differences
 $F_N^{(1)}{}'''$ and $F_N^{(2)}{}'''$ and confirmed that they quickly converge
 to the exact values of ${\rm vol}(M)$ up to  overall  factors $\frac{1}\pi$ and $ \frac{1}{4\pi^2}$ respectively, as we increase $N$. The results summarized in the table below show excellent agreement.  
 \begin{table}[h!]
    \begin{tabular}{ | c | c | cc | cc |}
    \hline
    $K$ & vol($S^3\backslash K$) & $  \pi F_N^{(1)}{}'''$ &  $(N)$   &  $4 \pi^2 F_N^{(2)}{}'''$ & $(N)$   \\ \hline
    $\mathbf{4}_1$& 2.02988   & 2.03001 &($27$)  & 2.02898  &($17$) \\ \hline
    $\mathbf{5}_2$& 2.82812  & 2.82828 &($12$) & 2.82674 &($12$)  \\ \hline
$\mathbf{6}_1$& 3.16396  & 3.20648 &($12$) & 3.15574 &($12$)  \\ \hline
$\mathbf{6}_2$& 4.40083  & 4.40364 &($12$) & 4.39929 &($12$)  \\ \hline
$\mathbf{6}_3$& 5.69302  & 5.69464 &($11$) & 5.68799 &($9$)  \\ \hline
$\mathbf{7}_2$& 3.33174  & 3.56613 &($12$) & 3.27455 &($12$)  \\ \hline
$\mathbf{7}_3$& 4.59213  & 4.58680 &($12$) & 4.58331 &($11$)  \\ \hline    
    \end{tabular}
    \label{table:4567}
\end{table}
The computation of 3-loop invariant $F_N^{(3)}$ takes significantly longer, due to the large
number of Feynman diagrams. We have done the computation for $\mathbf{4}_1$ 
and obtained $F_{N}^{(3)} =  0.03128, 0.02844, 0.02602$
for $N=7,8,9$.  It is thus strongly suggested that $\lim_{N\ra \infty} F_{N}^{(3)}/N^3=0$, 
in accordance with the holographic prediction.

{\it Discussion. }
In this Letter we have performed a quantitive study of $AdS_4/CFT_3$ arising from wrapped M5-branes, by comparing the free energy on both sides. 
We confirm the famous $N^3$-behavior of the M5-brane physics including an overall factor. It
is highly desirable to have an analytic proof of the predictions on the perturbative  $PGL(N)$ CS invariants  on hyperbolic 3-manifolds in the large $N$ limit. Studying other physical objects, such as defects, will certainly give new insights and deserve further exploration.

{\it Acknowledgments.}
We are grateful to J. P. Gauntlett and M. Yamazaki for invaluable comments on the manuscript.
This work was supported by  
NRF grants funded by the Korea government 
with grant No. 2010-0023121, 2012046278 (NK) and  
2012R1A1B3001085, 2012R1A2A2A02046739 (SL).  
\bibliography{w3c-2}
\end{document}